\documentclass[lettersize,journal]{IEEEtran}
\usepackage{amsmath,amsfonts}
\usepackage{algorithm}
\usepackage{algpseudocode}
\usepackage{array}
\usepackage{booktabs}
\usepackage[caption=false,font=normalsize,labelfont=sf,textfont=sf]{subfig}
\usepackage{textcomp}
\usepackage{stfloats}
\usepackage{url}
\usepackage{verbatim}
\usepackage{pifont}
\usepackage{graphicx}
\usepackage{cite}
\usepackage{tikz,xcolor}
\usepackage[implicit=false]{hyperref}

\hypersetup{hidelinks,
	colorlinks=true,
	allcolors=black,
	pdfstartview=Fit,
	breaklinks=true}

\begin{document}

\title{APT-ClaritySet: A Large-Scale, High-Fidelity Labeled Dataset for APT Malware with Alias Normalization and Graph-Based Deduplication}

\author{Zhenhao Yin, Hanbing Yan, Huishu Lu, Jing Xiong, Xiangyu Li, Rui Mei, Tianning Zang
\thanks{Hanbing Yan is the corresponding author: yanhanbing@iie.ac.cn}
\thanks{Zhenhao Yin is with School of Cyber Security, University of Chinese Academy of Sciences, Beijing 100049, China and also with Institute of Information Engineering, Chinese Academy of Sciences, Beijing 100085, China (email: yinzhenhao@iie.ac.cn)}
\thanks{Hanbing Yan and Tianning Zang are with Institute of Information Engineering, Chinese Academy of Sciences, Beijing 100085, China (email: \{yanhanbing, zangtianning\}@iie.ac.cn)}
\thanks{Huishu Lu, Jing Xiong and Xiangyu Li are with State Key Laboratory of Software Development Environment, Beihang University, Beijing 100191, China (email: \{lhs, xxx\_j, li\_xiangyu\}@buaa.edu.cn)}
\thanks{Rui Mei is with iFLYTEK Security Laboratory, Hefei, Anhui 230088, China and also with Software Security Research Group, Peking University, Beijing 100871, China (email: ruimei@pku.edu.cn)}% <-this % stops a space
}

% The paper headers
\markboth{Journal of \LaTeX\ Class Files,~Vol.~14, No.~8, August~2021}%
{Shell \MakeLowercase{\textit{et al.}}: A Sample Article Using IEEEtran.cls for IEEE Journals}

\IEEEpubid{0000--0000/00\$00.00~\copyright~2021 IEEE}

\maketitle

\begin{abstract}
Large-scale, standardized datasets for Advanced Persistent Threat (APT) research are scarce, and inconsistent actor aliases and redundant samples hinder reproducibility. This paper presents APT-ClaritySet and its construction pipeline that normalizes threat actor aliases (reconciling approximately 11.22\% of inconsistent names) and applies graph-feature deduplication—reducing the subset of statically analyzable executables by 47.55\% while retaining behaviorally distinct variants. APT-ClaritySet comprises: (i) APT-ClaritySet-Full, the complete pre-deduplication collection with 34{,}363 malware samples attributed to 305 APT groups (2006--early 2025); (ii) APT-ClaritySet-Unique, the deduplicated release with 25{,}923 unique samples spanning 303 groups and standardized attributions; and (iii) APT-ClaritySet-FuncReuse, a function-level resource that includes 324{,}538 function-reuse clusters (FRCs) enabling measurement of inter-/intra-group sharing, evolution, and tooling lineage. By releasing these components and detailing the alias normalization and scalable deduplication pipeline, this work provides a high-fidelity, reproducible foundation for quantitative studies of APT patterns, evolution, and attribution.
\end{abstract}

\begin{IEEEkeywords}
Advanced Persistent Threats, APT Malware Dataset, Alias Normalization, Graph-Based Deduplication
\end{IEEEkeywords}

\section{Introduction}

\IEEEPARstart{C}{omprehensive} understanding of the attack patterns, technological evolution, and organizational profiles of Advanced Persistent Threats (APTs) is crucial to strengthening cyber defense. High-quality, labeled APT malware datasets are foundational to such research. However, the APT research field \cite{krishnapriya2024comprehensive,hu2025high,myneni2020dapt,rahal2025dataset,rani2024comprehensive} faces a persistent shortage of large-scale, standardized, and high-quality datasets, which limits comprehensive analyses of APT behaviors \cite{rani2024comprehensive,salim2023systematic,myneni2020dapt,mabrouk2024lateral,articleindex1}.

Current efforts are hindered by four challenges. First, public APT datasets often lack sufficient sample size, organizational coverage, labeling accuracy, and timeliness, constraining machine learning and complex scenario analysis. Second, heterogeneous aliasing across vendors and institutions causes naming inconsistencies, impeding intelligence integration and comparative studies. Third, large malware corpora contain substantial near-duplicate samples, inflating storage and analysis costs and introducing statistical bias and overfitting risks. Finally, most datasets lack systematic function-level code reuse features, limiting insights into cross-group code sharing, technical provenance, and unique codebases.

To address these issues, this paper proposes a comprehensive methodology for constructing and characterizing a large-scale labeled APT malware dataset. We design an APT alias normalization system that maps heterogeneous vendor names to unified identifiers by integrating multi-source intelligence. We further develop an efficient graph-feature-based binary deduplication algorithm to remove functionally redundant yet slightly variant samples, improving dataset uniqueness and laying the groundwork for function-level analysis. We also build a function-level code reuse feature set to enable preliminary analysis of inter-/intra-group code reuse and support subsequent pattern mining. Based on these components, we construct a dataset family: APT-ClaritySet-Full (the labeled full corpus), APT-ClaritySet-Unique (the high-precision deduplicated subset), and APT-ClaritySet-Func (the function-level code reuse feature set).
\IEEEpubidadjcol
The main contributions are as follows. First, \textbf{Dataset Contribution}: We provide APT-ClaritySet-Full and its deduplicated subset APT-ClaritySet-Unique, comprising 25{,}923 unique samples attributed to 303 APT organizations, as a standardized benchmark. Second, \textbf{Methodology Contribution}: We establish an alias normalization system and workflow that resolves long-standing naming inconsistencies, improving label reliability and consistency. Third, \textbf{Algorithm Contribution}: We propose an efficient deduplication method based on graph features and a hybrid similarity metric, substantially reducing redundancy while maintaining high precision. Fourth, \textbf{Feature Engineering Contribution}: Through function-level feature extraction and clustering, we provide APT-ClaritySet-Func, enabling analysis of technical characteristics, code evolution, and inter-group relationships.

The remainder of the paper is organized as follows. Section~\ref{sec:background_motivation} motivates the problem and reviews limitations in existing resources. Section~\ref{sec:alias_normalization} details the alias normalization framework. Section~\ref{sec:deduplication} presents the graph-based deduplication method. Section~\ref{sec:dataset_desc} describes and evaluates APT-ClaritySet, including coverage, deduplication effectiveness, and function-level analysis. Section~\ref{sec:challenges} discusses challenges, advantages, and future directions. Section~\ref{sec:related_work} surveys related work, and Section~\ref{sec:conclusion} concludes.
\section{Background and Motivation}
\label{sec:background_motivation}
Advanced Persistent Threats (APTs) pose sustained, targeted risks to global cyberspace. Progress is constrained by the scarcity of high-quality, standardized, large-scale datasets, which limits reproducibility and the scope of measurement-driven research\cite{rani2024comprehensive,salim2023systematic,myneni2020dapt,mabrouk2024lateral,articleindex1}. This section concisely frames four gaps motivating this study: chaotic alias naming across threat actors, limitations of existing APT sample datasets, deduplication challenges at scale, and the absence of function-level code-reuse features.

\subsection{Chaotic APT Alias Naming}

The absence of a unified naming convention hinders precise identification and longitudinal tracking of APT organizations. Different vendors and institutes adopt divergent aliasing schemes\cite{urlindex2,urlindex3,urlindex4,urlindex5,urlindex6} (e.g., Microsoft's weather-based scheme\cite{microsoftactornaming} versus CrowdStrike's country-animal convention\cite{crowdstrikeactornaming}), producing the common ``one entity, multiple names'' phenomenon. This fragmentation complicates data alignment, cross-report comparison, and historical linkage, thereby degrading threat intelligence integration and attribution fidelity. Establishing a standardized alias mapping mechanism is thus essential.

\subsection{Existing APT Dataset Review}
High-quality, labeled APT malware datasets remain limited. For clarity, existing resources are grouped as: dedicated APT sample datasets, related malware datasets, and non-sample APT datasets.

\textbf{Dedicated APT sample datasets are scarce}: Public, APT-focused collections are few. APTMalware\cite{urlindex7} is small (3{,}500+ samples; 12 groups), while APTracker\cite{mazaheri2024aptracker} is larger (64{,}440 samples; 22 groups) but requires independent quality validation (Table~\ref{tab:apt_sample_datasets}).

To enable concrete comparison, concentration and diversity metrics are computed on public metadata of APTMalware, APTracker, and APT-ClaritySet: normalized Shannon entropy ($H_{\mathrm{norm}}$, higher is more diverse) and Herfindahl–Hirschman Index (HHI, lower is more diverse). For a categorical distribution $\{p_i\}_{i=1}^K$ with $\sum_i p_i=1$,
\begin{align}
H  = -\sum_{i=1}^{K} p_i \log p_i,\quad
H_{\mathrm{norm}}  = \frac{H}{\log K},\quad
\mathrm{HHI} = \sum_{i=1}^{K} p_i^{2}.
\end{align}

\begin{table*}[htbp]
\centering
\caption{Side-by-side comparison of dedicated APT malware datasets.}
\label{tab:apt_sample_datasets}
\scriptsize

\begin{tabular}{cm{1.5cm}<{\centering}m{1cm}<{\centering}m{1cm}<{\centering}m{0.7cm}<{\centering}m{1.7cm}<{\centering}m{1.9cm}<{\centering}m{4.5cm}<{\centering}}
\toprule
\textbf{Dataset Name} & \textbf{Collection Period} & \textbf{Claimed samples} & \textbf{Usable Samples} & \textbf{APT Groups} & \textbf{Org. Diversity} ($H_{\mathrm{norm}}$/HHI) & \textbf{Filetype Diversity} ($H_{\mathrm{norm}}$/HHI) & \textbf{Quality Control Findings} \\
\midrule
MVFCC\cite{haddadpajouh2020mvfcc} & 2019 or earlier & 1,200 & -- & 5 & -- & -- & Subset of \cite{urlindex7} \\
APTMalware\cite{urlindex7} & 2019 or earlier & 3,594 & 3,594 & 12 & \textbf{0.900} / 0.132 & 0.460 / 0.338 & Small scale; limited org coverage \\
\cmidrule(lr){1-8}
APTracker\cite{mazaheri2024aptracker} & 2020--2024$^*$ & 64,440 & 63,152 & 22 & 0.663 / 0.165 & 0.374 / 0.374 & 60 duplicate hashes and conflicting org labels; high org/type concentration \\
\cmidrule(lr){1-8}
APT-ClaritySet-Full & 2006--2025 & 34,363 & 34,363 & 305 & 0.784 / \textbf{0.020} & 0.579 / 0.186 & Alias normalization; hash dedup; manual spot checks \\
APT-ClaritySet-Unique & 2006--2025 & 25,923 & 25,923 & 303 & 0.7819 / 0.0211 & \textbf{0.6284 / 0.1459} & Graph-based deduplication \\
\bottomrule
\multicolumn{8}{l}{\footnotesize $^*$The authors' claimed collection window is used, not the samples' first-seen time; validation found some samples dating back to 2013.}\\
\multicolumn{8}{l}{\footnotesize APT-ClaritySet spans 2006--2025 and covers 305 APT groups.}
\end{tabular}
\normalsize
\end{table*}

The quantitative results are consistent with the above observations. For organization labels, APTracker exhibits $H_{\mathrm{norm}}=0.663$ and $\mathrm{HHI}=0.165$, indicating concentration in a few groups, whereas APT-ClaritySet attains lower concentration ($\mathrm{HHI}\approx0.020$) with broader, more balanced coverage. For file types, APT-ClaritySet achieves $H_{\mathrm{norm}}=0.579$ and $\mathrm{HHI}=0.186$, surpassing APTracker's 0.374/0.374, which is biased toward PE executables. Although APTracker reports 64{,}440 samples, 63{,}152 were accessible during independent validation, and 60 duplicate hashes with conflicting organization labels were identified, indicating quality-control gaps. Specifically for APT-ClaritySet-Unique, the organization-label diversity reaches $H_{\mathrm{norm}}=0.7819$ and $\mathrm{HHI}=0.0211$, while the file-type diversity reaches $H_{\mathrm{norm}}=0.6284$ and $\mathrm{HHI}=0.1459$, reflecting a more balanced post-deduplication composition compared to Full (0.579/0.186 on file types).

Comprehensive analysis reveals that existing APT datasets face systematic challenges: Dedicated datasets are extremely scarce, with only 2 specialized APT datasets of varying quality; Scale and timeliness contradictions, where large-scale datasets are outdated while recent datasets are limited in scale; Inconsistent labeling systems, with chaotic APT group naming severely hindering data integration; Absent quality control mechanisms, lacking systematic deduplication and validation processes; Unbalanced coverage, with significant biases in APT group, attack type, and temporal distributions. These systematic limitations highlight the urgent need for constructing large-scale, high-quality, standardized APT datasets.

\subsection{Deduplication Challenges and Needs}
Large-scale malware corpora often contain redundant or near-duplicate samples (e.g., variants or recompilations), inflating analysis cost and skewing statistics. Exact hashing (e.g., MD5) fails on code variants\cite{botacin2021understanding, cesare2010fast,or2019dynamic}. Fuzzy hashing (e.g., ssdeep) faces precision/efficiency limits on complex variants and incurs $O(n^2)$ comparisons at scale\cite{kida2022nation,naik2021fuzzy}. APT-oriented deduplication therefore requires high precision, scalability, and the ability to retain semantically meaningful variants while removing purely redundant ones.

\subsection{Missing Binary Function Reuse Clusters}
In this paper, we define binary function reuse clusters as sets of code-equivalent or near-duplicate functions aggregated across samples and groups.

Functions are the fundamental units of code functionality, and analyzing their reuse patterns is crucial for uncovering code sharing, technology adoption, and evolutionary paths among APT organizations\cite{basnet2024advanced,tereszkowski2024study}. However, existing APT datasets generally lack systematic function-level code reuse features, primarily due to the technical challenges of function-level analysis (e.g., function boundary identification, normalization, and large-scale similarity computation). The absence of such features limits the fine-grained comparison and tracing of APT toolkits and technical characteristics.

In sum, alias-name fragmentation, the paucity of high-quality APT corpora, technical bottlenecks in scalable deduplication, and the lack of function-level reuse features collectively hinder rigorous APT measurement. This study addresses these gaps by constructing APT-ClaritySet and its accompanying methodologies, providing a standardized, high-quality foundation for the research community.

\section{Dataset Construction Pipeline}

This section elaborates on the systematic methodology and pipeline for constructing APT-ClaritySet, a large-scale, high-quality, labeled APT malware dataset. It aims to address core challenges in APT research, including data acquisition, label normalization, sample redundancy, and deep feature extraction.

\subsection{Overview}

\begin{figure*}[h!]
    \centering
    \includegraphics[width=1.0\textwidth]{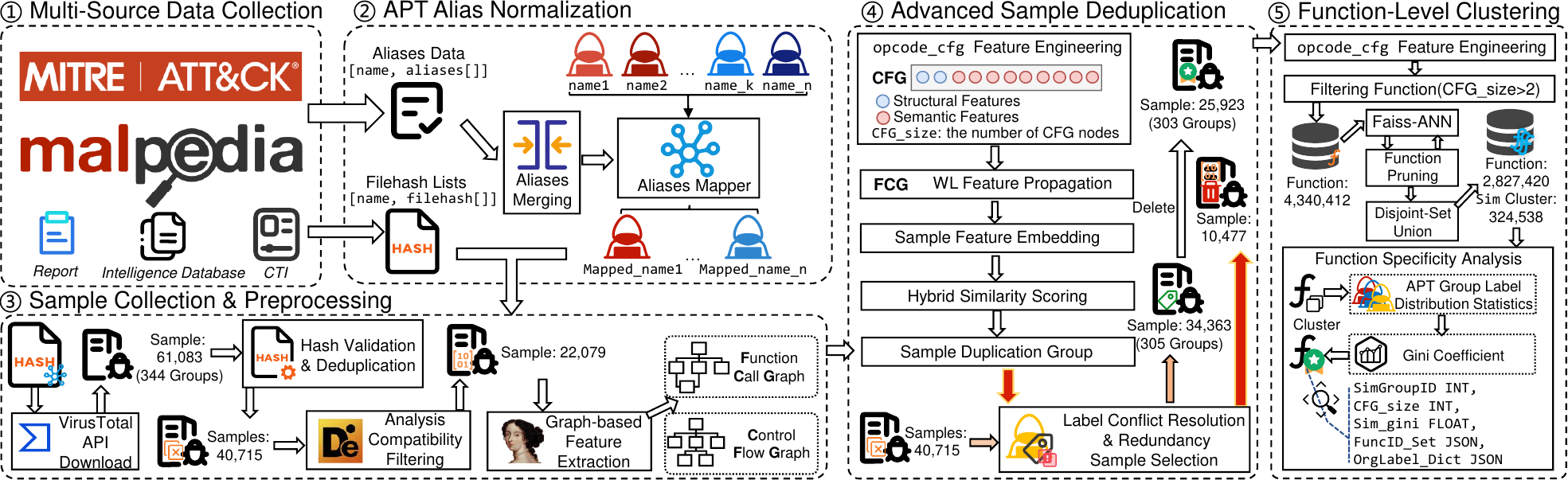}
    \caption{Overview of the APT-ClaritySet Construction Methodology}
    \label{fig:overview}
\end{figure*}

APT-ClaritySet is built through an iterative pipeline (Fig.~\ref{fig:overview}): (i) multi-source collection and integration; (ii) initial label extraction with APT alias normalization; (iii) data cleaning with integrity checks, exact-hash deduplication, label consistency assessment, and analyzability filtering; (iv) advanced feature-based deduplication; (v) multi-level annotation and quality control; and (vi) function-level feature extraction and clustering. The design targets dataset scale, label fidelity, sample uniqueness, and feature richness.

\subsection{Multi-Source Data Collection}

To ensure the comprehensiveness and representativeness of the dataset, a multi-source acquisition strategy was adopted. First, via an anonymized threat intelligence aggregation platform, APT sample hashes, preliminary organizational attributions, and links to original sources (e.g., vendor blogs and CTI reports) were collected. Second, given the hashes, binaries and enriched metadata were retrieved from VirusTotal (VT). Following this process, 61{,}083 labeled hashes were consolidated prior to cleaning.

\subsection{Initial Labeling and Alias Normalization}

Initial APT attributions were derived from sample metadata, associated reports, and VT multi-engine results. Due to inconsistent vendor naming (Section~\ref{sec:background_motivation}), the alias normalization system in Section~\ref{sec:alias_normalization} was applied to map heterogeneous names to unified primary identifiers, improving label consistency. Approximately 11.22\% of attributions that would otherwise be split were unified (details in Section~\ref{subsec:alias_normalization_effect}).

\subsection{Data Cleaning and Preprocessing}

After obtaining the initial set and normalized labels, a rigorous pipeline with explicit statistics was executed:
\begin{enumerate}
\item \textbf{Integrity Check and Exact-Hash Deduplication}: SHA-256 was recalculated locally for all files; corrupted/incomplete files were removed, and byte-identical files were collapsed. Conflicting labels within an exact-hash group were flagged. The set reduced from 61{,}083 to 40{,}715.
\item  \textbf{Label Consistency Assessment}: For groups with conflicting attributions, the Gini coefficient ($G = 1 - \sum_{i=1}^{n} p_i^2$) was used; when $G \le 0.2$, the majority label was retained; otherwise, labels were set to ``unknown''. About 15.6\% of disputed cases were processed.
\item  \textbf{Analyzability Filtering}: Executable type (PE/ELF) and packing state (DIE\cite{detect-it-easy}, VirusTotal\cite{virustotal}) were screened, prioritizing unpacked or successfully unpacked binaries, yielding 22{,}079 analyzable binaries.
\end{enumerate}
\subsection{Advanced Deduplication}
\label{subsec:AdvDedup}
Preliminary exact-hash deduplication cannot collapse near-duplicates that are functionally equivalent but differ slightly in code (e.g., variants only replacing hard-coded Indicators of Compromise, IOCs). To address such cases, an efficient graph-feature-based deduplication (Section~\ref{sec:deduplication}) was applied: structural features from function-level CFGs and semantic features from opcode sequences were combined into an \texttt{opcode\_cfg} representation; a hybrid similarity was computed (Gaussian-kernelized Euclidean for structure, cosine for semantics); a strict threshold (e.g., 0.9999, calibrated on held-out pairs; Appendix~\ref{append:Ablation}) and DSU were used to merge near-duplicate clusters with high precision. On CentOS~8.5 with an Intel Xeon Silver~4210R and 128GB RAM, feature extraction succeeded for 22{,}035 of 22{,}079 binaries (44 parsing/timeouts), and deduplication over these 22{,}035 binaries took 631.55s on average, removing 10{,}477 near-duplicates (47.55\%).

\subsection{Function Feature Extraction and Clustering}

\begin{figure}[ht!]
  \centering
  \includegraphics[width=0.45\textwidth]{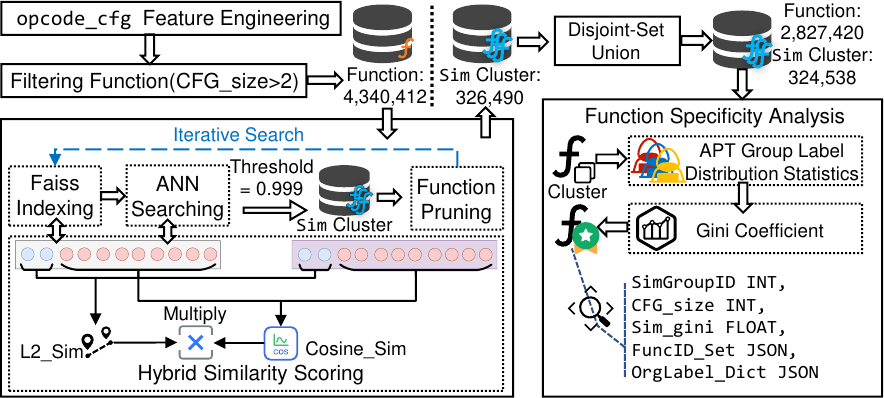}
  \caption{Overview of the Function-Level Clustering}
  \label{fig:function_clustering_flowchart}
\end{figure}

To enable downstream measurement of code reuse, a scalable function-level feature extraction and clustering workflow was employed (Fig.~\ref{fig:function_clustering_flowchart}). The pipeline is engineering-oriented and \emph{pluggable}, allowing alternative function similarity backends from the literature without affecting downstream measurements.

\textbf{Function-Level \texttt{opcode\_cfg} Feature Extraction}: All functions were extracted, and trivial ones with overly simple CFGs (\emph{$\le 2$ basic blocks}) were filtered. Such functions are mainly import thunks, trampolines, or thin wrappers with limited discriminative value; excluding them reduces noise and improves precision with negligible loss in reuse discovery (Appendix~\ref{append:Ablation}). For retained functions, structural vectors $V_{struct,f}$ and semantic vectors $V_{sem,f}$ were computed per Section~\ref{sec:deduplication}, followed by L2-normalization of $V_{sem,f}$.

\textbf{Large-Scale Function Similarity Clustering}: A hybrid similarity was used:
\begin{equation}
Sim_{struct}(f_i, f_j) = \exp(-\gamma \cdot d_{L2}(V_{struct,i}, V_{struct,j})^2)
\end{equation}
\begin{equation}
Sim_{sem}(f_i, f_j) = \hat{V}_{sem,i} \cdot \hat{V}_{sem,j}
\end{equation}
\begin{equation}
Sim_{hybrid} = Sim_{struct} \cdot Sim_{sem}.
\end{equation}
An iterative Search-and-Prune strategy with Faiss\cite{douze2024faiss} ANN\cite{approximate-nearest-neighbors} generated candidates; exact $Sim_{hybrid}$ was then computed and DSU applied under a high-precision threshold (e.g., 0.999).

\textbf{Function Reuse Clusters and Organizational Distributions}: Clusters containing functions from multiple organizations were treated as reuse clusters. For each, per-organization distributions were computed to reveal cross-organization sharing and group-preferred functions; cluster IDs, sizes, distributions, and representative metadata were stored for downstream measurement. While the technique is generally applicable to malware, its usage here is APT-centric through normalized attributions.

\section{APT Alias Normalization}
\label{sec:alias_normalization}

The precise identification of Advanced Persistent Threat (APT) organizations is hindered by the absence of a unified naming convention across vendors and platforms. Disparate aliases for the same entity lead to fragmented intelligence and hinder reproducible measurement. We therefore design a systematic alias mapping and normalization framework that integrates heterogeneous sources into a dynamically updated, queryable knowledge base, establishing reliable organizational labels for large-scale sample analysis.
 
\subsection{Collection and Integration}

We compiled aliases from multiple authoritative and public intelligence sources:
\begin{itemize}
\item \textbf{Vendor reports and blogs}: FireEye/Mandiant, CrowdStrike, Kaspersky, Symantec/Broadcom, among others.
\item \textbf{Open intelligence repositories}: MITRE ATT\&CK\textsuperscript{®}\cite{mitreattack}, Malpedia\cite{malpedia}.
\item \textbf{Academic and industry studies}: peer-reviewed publications and conference/industry reports.
\end{itemize}
A semi-automated pipeline first filtered irrelevant entries; researchers then verified extracted name pairs. In total, 1{,}553 raw name/alias entries referring to about 960 potential APT entities were collected from 200+ sources, forming the initial corpus for normalization.

\subsection{Normalization Mapping Rules}

\begin{figure}[h!]
  \centering
  \includegraphics[width=0.4\textwidth]{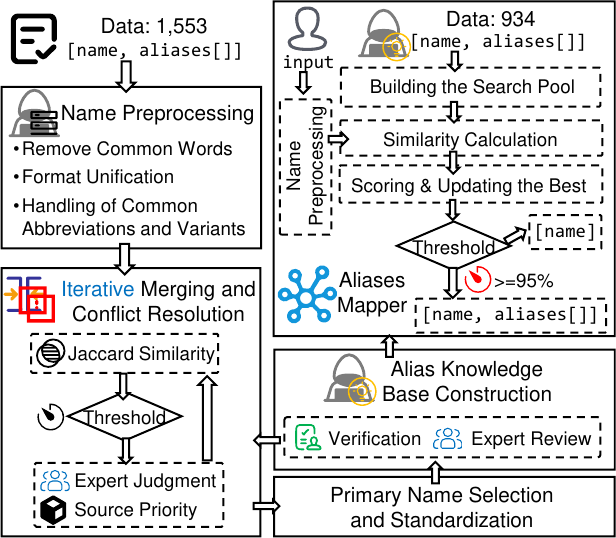} 
  \caption{Workflow of the APT Alias Normalization and Mapping System}
  \label{fig:alias_normalization_flowchart}
\end{figure}

Our goal is to maximize alias coverage while ensuring mapping accuracy. The workflow is summarized in Fig.~\ref{fig:alias_normalization_flowchart}.
\begin{enumerate}
\item \textbf{Preprocessing}: remove non-specific modifiers (e.g., ``team'', ``group''), unify format (case-folding, Unicode normalization, punctuation/space normalization), and normalize common abbreviations/variants.
\item \textbf{Merging and conflict resolution}: iteratively merge entity records using alias-set similarity (e.g., Jaccard above a tuned threshold) and source authority weighting; ambiguous cases are escalated for expert arbitration.
\item \textbf{Primary name standardization}: select a canonical name based on industry prevalence, first-use provenance, and recommendations from standardized frameworks (e.g., MITRE ATT\&CK\textsuperscript{®}).
\end{enumerate}

\subsection{Construction and Validation}

The system comprises a structured knowledge base (JSON) and a query interface. Inputs are preprocessed and matched using Levenshtein (edit-based) and Token Sort Ratio (token-based) similarity. Matches with similarity $\geq$95\% are accepted; 80–95\% are candidates for analyst review. Results return the canonical name, known aliases, and source attribution.

To assess reliability, three independent analysts ($>5$ years of threat intelligence experience) reviewed a representative subset, focusing on $\sim$15\% low-confidence or conflicting mappings. Each analyst independently judged correctness; disagreements were resolved by consensus. Inter-analyst agreement, measured by Cohen's $\kappa$~\cite{cohen1960coefficient}, was 0.93, evidencing the robustness of the normalization. The review prompted minor adjustments (approximately 5\%) and informed threshold/rule refinements. The knowledge base is scheduled for quarterly updates with expert screening.

\subsection{Impact on Dataset Consistency}
\label{subsec:alias_normalization_effect}

The normalization substantially improved label consistency:
\begin{itemize}
\item \textbf{Label unification}: Among 401 records with alias-induced inconsistencies, 45 (11.22\%) were unified under canonical APT names.
\item \textbf{Improved correlation}: In the final 25{,}923-sample corpus, 9{,}367 samples (36.13\%) were correctly consolidated under a single standardized entity, mitigating misattribution from naming discrepancies.
\end{itemize}
These results demonstrate that alias normalization effectively reduces fragmentation and strengthens the reliability of large-scale APT measurement and attribution.

\section{Efficient Graph-Based Binary Deduplication}
\label{sec:deduplication}

Redundant samples are prevalent in large malware datasets, increasing storage and computational costs, leading to biases in statistical analysis, and affecting the accuracy of deeper analyses such as code reuse. Accurate deduplication is a prerequisite for ensuring the precision of analytical results. An efficient graph-feature-based deduplication method is presented to identify redundant samples that are functionally identical despite minor code variations.

\subsection{\texttt{opcode\_cfg} Feature Design}

\begin{figure}[h!]
\centering
\includegraphics[width=0.47\textwidth]{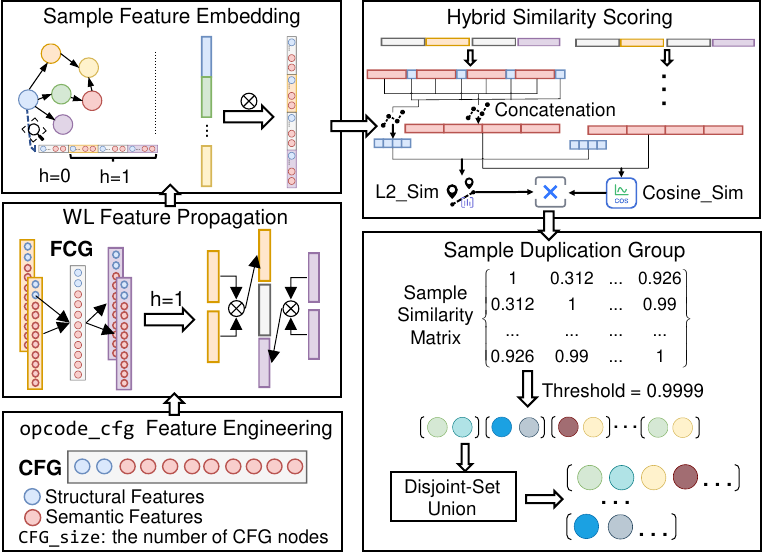}
\caption{Overall Workflow of the Graph-Feature-Based Binary Sample Deduplication Method}
\label{fig:deduplication_pipeline}
\end{figure}

\textbf{Method Overview}. This method (Figure~\ref{fig:deduplication_pipeline}) first extracts Control Flow Graphs (CFGs) and Function Call Graphs (FCGs), designs an \texttt{opcode\_cfg} feature that fuses structural and semantic information, employs the Weisfeiler-Lehman (WL) algorithm to enhance node representations, and finally completes deduplication through a hybrid similarity metric.

To characterize functional similarity, CFGs, FCGs, and assembly are extracted using IDA Pro\cite{idapro}. An \texttt{opcode\_cfg} feature is designed to integrate CFG structural information and opcode-sequence semantics.

\textbf{\texttt{opcode\_cfg} Feature Definition and Computation}. The \texttt{opcode\_cfg} feature aims to fuse function control flow structure with instruction-level semantics. For a function $f$'s CFG $G=(V, E)$, each basic block $b \in V$ has a node ID $id_b$, number of instructions $n_b$, out-degree $deg_b^+$, and occurrence counts $f_{C_k,b}$ for various opcode categories $C_k$. The \texttt{opcode\_cfg} feature vector $F_{opcode\_cfg}$ is 11-dimensional:

\begin{equation}
F_{opcode\_cfg} = [X, Y, W_{C_1}, W_{C_2}, \ldots, W_{C_9}]
\end{equation}

It contains two types of information:

\textbf{Structural Features}: $X$ (weighted node ID) and $Y$ (weighted node out-degree) reflect control flow distribution and branching complexity, drawing inspiration from Chen et al.'s Centroid method \cite{chen2014achieving}:
\begin{align}
\omega &= \sum_{(s,d) \in E} (n_s + n_d) \label{eq:omega} \\
X &= \frac{\sum_{(s,d) \in E} (n_s \cdot id_s + n_d \cdot id_d)}{\omega} + C_{offset} \label{eq:weighted_node_id} \\
Y &= \frac{\sum_{(s,d) \in E} (n_s \cdot deg_s^+ + n_d \cdot deg_d^+)}{\omega} \label{eq:weighted_out_degree}
\end{align}
    
where $n_s, n_d$ are the number of instructions in basic blocks, $id_s, id_d$ are node IDs, $deg_s^+, deg_d^+$ are out-degrees, $\omega$ is a normalization factor, and $C_{offset}$ is an offset correction value. The $Z$ feature (weighted loop depth) was initially considered and defined as:
$Z = \frac{\sum_{(s,d) \in E} (n_s \cdot ldt_s + n_d \cdot ldt_d)}{\omega}$, 
where $ldt_s, ldt_d$ are the loop depths of basic blocks $s$ and $d$. However, $Z$ was discarded due to high computational cost, path explosion issues, and no significant performance improvement (see Section~\ref{subsec:ablationstudy}, RQ3). The $X$ and $Y$ features aim to capture the macroscopic morphology of control flow and exhibit some robustness to local code variations.

\textbf{Semantic Features}: $W_{C_1}, \ldots, W_{C_9}$ represent the weighted occurrence frequencies of 9 opcode categories along control flow paths. Opcodes are categorized into 9 semantic classes (MEM\_ACCESS, ARITHMETIC\_LOGIC, CONTROL\_FLOW, LOAD\_CONSTANT, OBJECT\_ORIENTED, TYPE\_CONVERSION, STACK\_OPS, METADATA\_REFLECTION, and OTHER\_IGNORE). For category $C_k$, its weight $W_{C_k}$ is:
    \begin{equation}
    W_{C_k} = \sum_{(s,d) \in E} (f_{C_k,s} + f_{C_k,d})
    \end{equation}
    This reflects the activity level of that opcode category along control flow paths.

\textbf{Opcode taxonomy}: Opcodes are classified into nine semantic categories according to their functional roles.
\begin{itemize}
    \item MEM\_ACCESS: explicit memory read/write or effective-address computation.
    \item ARITHMETIC\_LOGIC: arithmetic/bitwise operations and shifts on register/stack values (no explicit memory operand).
    \item CONTROL\_FLOW: direct/indirect unconditional/conditional transfers, calls/returns, multi-way branches, exception-related transfers.
    \item LOAD\_CONSTANT: materializing immediates/constants; for managed code, constant-pool and literal loads.
    \item OBJECT\_ORIENTED: object/array allocation, virtual/interface dispatch, type checks, synchronization (managed runtimes).
    \item TYPE\_CONVERSION: sign/zero extension, width/representation casts, and reinterpretation.
    \item STACK\_OPS: push/pop, frame setup/teardown, SP arithmetic, and managed local/argument load/store.
    \item METADATA\_REFLECTION: metadata-token loads and reflective invocation (managed runtimes).
    \item OTHER\_IGNORE: non-semantic nops/padding/debug traps/hints (down-weighted or ignored).
\end{itemize}

\textbf{Weisfeiler-Lehman Feature Propagation}. To capture richer graph structures, the WL algorithm iteratively aggregates neighborhood features to enhance node representations: (1) initialize each node with its 11-dimensional \texttt{opcode\_cfg} vector; (2) iteratively aggregate with neighbors; (3) aggregate final node features into a graph-level vector $\mathbf{G}$.

\textbf{Feature Representation Method}. A numerical representation is adopted for \texttt{opcode\_cfg}, which is less sensitive to minor variations than string hashing and enables fine-grained measurement and precise thresholding.

\subsection{Hybrid Similarity Metric}

This section proposes a hybrid similarity metric method based on the \texttt{opcode\_cfg} features and node-level enhanced features processed by WL. The enhanced feature $\mathbf{f}_v$ of a node $v$ is formed by concatenating multiple 11-dimensional \texttt{opcode\_cfg} feature blocks $\mathbf{b}_{v,k}$. Algorithm~\ref{alg:hybrid_similarity} describes the calculation process.

\begin{algorithm}
\caption{Binary Sample Hybrid Similarity Calculation}
\label{alg:hybrid_similarity}
\begin{algorithmic}[1]
\Require Binary samples $i, j$; Enhanced node features $\{\mathbf{f}_v^{(i)}\}_{v \in V_i}$, $\{\mathbf{f}_v^{(j)}\}_{v \in V_j}$; Gaussian kernel parameter $\gamma$.
\Ensure Hybrid similarity score $S_{\text{hybrid}}(i,j)$.

\State \textit{// 1. Global Feature Vector Construction}
\Function{ConstructGlobalVectors}{sample $x$}
    \State Initialize $\mathbf{V}_{\text{sem},x} \gets \emptyset$, $\mathbf{V}_{\text{struct},x} \gets \emptyset$
    \ForAll{node $v$ in sample $x$}
        \ForAll{block $\mathbf{b}_{v,k}$ in $\mathbf{f}_v^{(x)}$}
            \State Extract $(X_{v,k}, Y_{v,k})$, $(W_{C_1,v,k}, \ldots, W_{C_9,v,k})$ from $\mathbf{b}_{v,k}$
            \State $\mathbf{V}_{\text{sem},x} \gets \mathbf{V}_{\text{sem},x} \oplus [W_{C_1,v,k}, \ldots, W_{C_9,v,k}]$
            \State $\mathbf{V}_{\text{struct},x} \gets \mathbf{V}_{\text{struct},x} \oplus [X_{v,k}, Y_{v,k}]$
        \EndFor
    \EndFor
    \State \Return $\mathbf{V}_{\text{sem},x}, \mathbf{V}_{\text{struct},x}$
\EndFunction

\State

\State $(\mathbf{V}_{\text{sem},i}, \mathbf{V}_{\text{struct},i}) \gets \text{ConstructGlobalVectors}(i)$
\State $(\mathbf{V}_{\text{sem},j}, \mathbf{V}_{\text{struct},j}) \gets \text{ConstructGlobalVectors}(j)$

\State

\State \textit{// 2. Semantic Similarity ($S_1$)}
\State $S_1(i,j) \gets \text{cosine\_sim}(\mathbf{V}_{\text{sem},i}, \mathbf{V}_{\text{sem},j})$

\State

\State \textit{// 3. Structural Similarity ($S_2$)}
\State $\mathbf{V}'_{\text{struct},i} \gets \text{L2\_normalize}(\mathbf{V}_{\text{struct},i})$
\State $\mathbf{V}'_{\text{struct},j} \gets \text{L2\_normalize}(\mathbf{V}_{\text{struct},j})$
\State $S_2(i,j) \gets \exp(-\gamma \cdot \|\mathbf{V}'_{\text{struct},i} - \mathbf{V}'_{\text{struct},j}\|_2^2)$

\State

\State \textit{// 4. Hybrid Similarity}
\State $S_{\text{hybrid}}(i,j) \gets S_1(i,j) \cdot S_2(i,j)$

\State \Return $S_{\text{hybrid}}(i,j)$

\end{algorithmic}
\end{algorithm}

The calculated $S_{\text{hybrid}}(i,j)$ quantifies sample similarity. If $S_{\text{hybrid}}(i, j) \ge \tau$ (e.g., 0.9999), they are considered a duplicate sample pair.
\textit{Threshold selection}. $\tau$ is selected on a held-out validation set by maximizing F1 under a zero false-merge constraint (precision=1.0). The optimal region concentrates near $\tau\in[0.9997,0.9999]$; $\tau=0.9999$ achieved the best validation F1 and remained robust under $\pm2\times10^{-4}$ perturbations.

\subsection{Ablation Study and Comparison}
\label{subsec:ablationstudy}
\noindent\textit{Ground-truth duplicate construction}. ``Duplicate'' denotes functionally equivalent, highly similar binaries rather than byte-identical files. To assemble the ground-truth set (221 duplicate groups; 1181 samples), original malware binaries were minimally mutated while preserving functionality, including: (i) instruction-preserving edits (register renaming, NOP padding, basic-block reordering without changing control dependencies); (ii) non-functional constant/resource edits (hard-coded string tweaks, timestamp/metadata changes); and (iii) benign refactoring of non-critical helper routines. Non-identity was enforced by distinct cryptographic hashes; functional equivalence was confirmed via sandbox execution (consistent process tree, file/network I/O, exit behavior), stable interface checks, and manual differential analysis on a sampled subset.

To validate component effectiveness and determine the optimal configuration, ablation studies were conducted on a ground-truth test set containing 221 duplicate sample groups (1181 samples). We evaluated the impact of each component on the F1-score, focusing on 5 research questions (RQs) (details in Appendix~\ref{append:Ablation}).

\begin{enumerate}
\item \textbf{RQ1: Numerical Representation Effectiveness}—Numerical representation outperforms string hashing with F1 improvements exceeding 0.05.
\item \textbf{RQ2: Opcode Features}—Introducing semantic features ($W_{C_k}$) significantly enhances discriminative power, improving F1.
\item \textbf{RQ3: \texttt{CFG\_size} Weighting}—Applying \texttt{CFG\_size} weighting to structural features ($X$, $Y$) better reflects function complexity and significantly improves performance (F1 increased by nearly 0.04).
\item \textbf{RQ4: Necessity of Z Feature}—The $Z$ feature is computationally expensive and did not yield significant gains; it is discarded.
\item \textbf{RQ5: Comparison of Distance Metrics}—The hybrid metric (combining structural and semantic similarity) provides more reliable judgments, with an F1 improvement exceeding 0.03.
\end{enumerate}

Experiments (Appendix~\ref{sec:AblationResult}) demonstrate that numerical representation, the hybrid metric, and CFG weighting contribute significantly. The final recommended configuration is: $X,Y$ structural features (without $Z$), opcode semantic features, \texttt{CFG\_size} weighting, and the hybrid similarity metric. This configuration achieved an F1$\approx$1 on the test set (threshold 0.9999), effectively identifying functionally similar samples, and provides the basis for the advanced deduplication in Section~\ref{subsec:AdvDedup}.

\section{APT-ClaritySet: Description and Evaluation}

\label{sec:dataset_desc}

This section aims to comprehensively describe the core characteristics of the large-scale labeled APT malware dataset, APT-ClaritySet, constructed in this study, and to systematically evaluate its quality. By providing a detailed description of the dataset's scale, coverage, content composition, and included derived features, combined with multi-dimensional quality assessment results, we intend to demonstrate the reliability, value, and potential contributions of this dataset to the APT research community. Concurrently, this section will also candidly discuss the dataset's existing limitations and potential biases, offering necessary background information for researchers utilizing this dataset.

\subsection{Publicly Released Dataset Components}
This research will publicly release the following three core dataset components to support related research in the APT domain. All components adopt standardized APT group attributions through the alias-normalization pipeline (Section~\ref{sec:alias_normalization}), reconciling approximately 11.22\% of inconsistent aliases across sources to improve reproducibility and longitudinal analyses.

\begin{figure}[htbp!]
\centering
\includegraphics[width=0.45\textwidth]{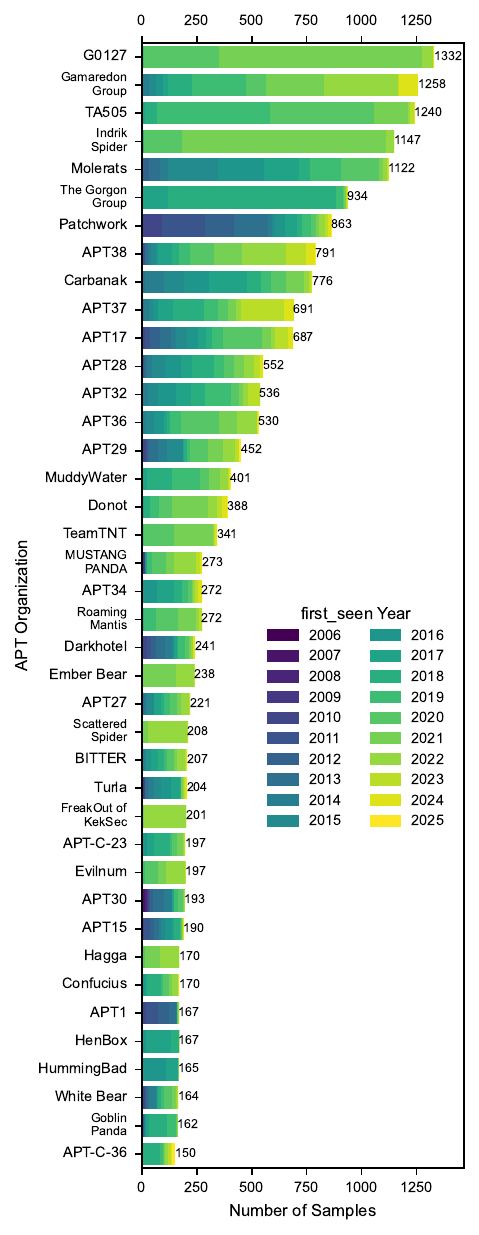}
\caption{Temporal Distribution of Major APT Groups by Sample Count}
\label{fig:aptsamplebyyear}
\end{figure}

\begin{enumerate}
\item \textbf{APT-ClaritySet-Full (pre-deduplication)}: Contains 34,363 malware samples associated with 305 APT groups (collection window: 2006--early 2025), prior to advanced deduplication. This near-original collection supports studies on deduplication efficacy and scenario-specific analyses while preserving standardized attributions.
\item \textbf{APT-ClaritySet-Unique (deduplicated)}: The primary release produced by the graph-feature deduplication method proposed in this study, containing 25,923 unique samples covering \textbf{303} APT groups. It offers higher uniqueness and analytical value, and is the main subject in subsequent analyses. The temporal distribution of major APT groups (those with $\geq 150$ samples) is shown in Figure~\ref{fig:aptsamplebyyear}, reflecting both activity evolution and sample volume distribution across groups.
\item \textbf{APT-ClaritySet-FuncReuse (function-level)}: A function-level resource comprising 324,538 function-reuse clusters (FRCs) extracted from a subset of samples in APT-ClaritySet. Each FRC includes a unique identifier, cluster size (number of functions), distribution across associated APT groups, and representative-function metadata (e.g., function start address in the original sample, sample hash, structured Control Flow Graph (CFG), and a simplified assembly-code summary) to enable research on code provenance and sharing patterns.
\end{enumerate}

\subsection{Dataset Core Characteristics Overview}

\textbf{Scale and Coverage:} APT-ClaritySet-Unique comprises 25,923 valid malware samples, covering 303 APT groups. The compilation timestamps primarily range from 2004 to 2023, with first-seen dates recorded from 2006 to early 2025. The peak activity period is concentrated between 2018 and 2022, with targets spanning multiple regions and industries.

\textbf{Content Features:} The dataset includes diverse file types, primarily: executable files (approximately 51.6\%, mainly Windows PE), document-based payloads (approximately 24.6\%), script files (approximately 7.0\%), mobile platform samples (mainly Android, approximately 7.7\%), and archive/container formats (approximately 4.7\%). Regarding basic statistical features, the average sample size is 1.817MB (median 234.55KB), and the average file entropy is 6.49. Among analyzable PE and ELF files, approximately 15.2\% were identified as packed.

\subsection{Deduplication Effectiveness}

To construct the high-fidelity core dataset, we systematically deduplicated APT-ClaritySet-Full (34,363 samples, 305 APT groups). Applying our graph-feature method to 22,035 statically analyzable executables removed 10,477 functionally similar redundant samples, reducing this analyzable subset by 47.55\% while preserving behaviorally distinct variants. The resulting APT-ClaritySet-Unique contains 25,923 unique malware samples spanning 303 APT groups, a net reduction of 8,440 samples (approximately 24.56\%) compared with APT-ClaritySet-Full, thereby improving uniqueness and analytical efficiency.

\subsection{Function-Level Code Reuse Analysis}

To facilitate research on code provenance and sharing, we construct and publicly release a unction-level code reuse feature set comprising over 4.3 million functions extracted from 9,512 analyzable samples within APT-ClaritySet and clustered into 324,538 function-reuse clusters (FRCs).

Each FRC captures a set of functions with highly similar code structures and opcode semantics, and provides a unique identifier, cluster size, distribution across associated APT groups, and representative-function metadata (function start address, sample hash, structured CFG, and simplified assembly-code summary). These features enable quantification of inter-/intra-group code sharing, family evolution analysis, and the discovery of tooling lineage and technology transfer.

\textbf{Distribution Characteristics of APT Group Labels in FRCs}: The dataset supports analysis of FRC dispersion across APT groups, aiding identification of group-preferred code snippets.

\subsection{Dataset Quality Assessment}

We evaluate dataset quality along four dimensions—completeness, accuracy, representativeness, and usability—combining automated statistics, stratified expert intelligence-based attribution, and comparisons with public benchmarks.

\textbf{Completeness}: Core attribute fields (hash value, file type, group label, first-seen time, file entropy) have a fill rate of 100.0\%.

\textbf{Accuracy}: For reliable APT group attribution, label accuracy was evaluated on the APT-ClaritySet-Full (34,363 samples, 305 APT groups) \textit{prior} to deduplication to avoid circularity, as the deduplication algorithm leverages original group labels (OrgLabel). Using optimal stratified random sampling and expert intelligence-based verification (Appendix~\ref{appendix:sampling}), 1,906 samples (5.54\%) covering all 305 groups were assessed. Among them, 79.1\% had original intelligence records retrieved directly from repositories; 20.9\% were verified via external attribution. The proportions of correct, suspicious, incorrect, and unknown labels were 96.43\%, 2.10\%, 0.58\%, and 0.89\%, respectively. Thus, the accuracy for high-quality labels (``correct'') is 96.43\% with a 95\% Clopper--Pearson interval of [95.50\%, 97.22\%]; counting ``correct''+``suspicious'' as usable yields 98.53\% with [97.88\%, 99.02\%]. Fleiss' Kappa among reviewers reached 0.94, indicating high inter-rater reliability. Accordingly, at a 95\% confidence level, the estimated attribution-label accuracy is 96.43\%.

\textbf{Representativeness}: Regarding APT group coverage, the dataset includes 303 groups. Against a comprehensive benchmark integrating and deduplicating multiple public APT lists (MITRE ATT\&CK\textsuperscript{®}, Malpedia, and others) using our alias normalization system (Section~\ref{sec:alias_normalization}), we identified 934 unique APT entities; thus, coverage is approximately 32.44\% (303/934). The included 303 groups encompass the majority of persistently active, well-documented, and high-impact actors in the global landscape, whereas many excluded entities exhibit minimal activity, sparse public intelligence, or discontinued operations, complicating large-scale acquisition and validation. Sample diversity is reflected across file types and claimed target industries, and the temporal first-seen distribution aligns with public reports on global APT trends.

\textbf{Usability}: Metadata and features are provided in structured formats (e.g., JSON Lines or CSV). Core metadata and basic static features achieve 100\% extraction; for supported sample types, function-level code reuse features achieve complete coverage. Documentation includes field descriptions, a methodology overview, and access guidelines to facilitate academic use.

\subsection{Security and Ethics}

This research adheres to security-research ethical guidelines to ensure a secure, compliant process with effective risk management.
\begin{itemize}
\item \textbf{Sample Handling Security}: Samples are processed and analyzed in physically isolated, virtualized secure environments; stored encrypted on restricted-access servers with encrypted transmission; automated processing is prioritized, and dynamic analysis is conducted in secure sandboxes.
\item \textbf{Data Sharing and Access Control}: Sharing of raw samples requires a strict Data Use Agreement (DUA), limited to academic research, prohibiting malicious use or re-dissemination. All shared raw samples are ``disarmed'' before provision—e.g., by removing direct execution capabilities—to minimize accidental execution risk while preserving static-analysis value.
\item \textbf{Potential Risks and Mitigation}: Leakage risks are mitigated via isolation, encryption, and access control. Misuse is prevented through usage protocols and limited disclosure of sensitive technical details. The impact of attribution error is reduced via multi-source cross-validation and explicit reporting of label confidence and accuracy.
\item \textbf{Research Value and Social Responsibility}: The dataset aims to enhance understanding of APT attacks, improve cyber defense capabilities, and contribute to objective scientific knowledge, advocating transparency and practical guidance.
\end{itemize}

\subsection{Limitations and Biases}

Despite efforts to construct a high-quality dataset, several limitations and potential biases remain:
\begin{itemize}
\item \textbf{Attribution and Visibility Bias}: Reliance on public intelligence may favor active or easily detectable groups, underrepresenting covert actors.
\item \textbf{Historical Detection Constraints}: Early APT samples are relatively scarce and may not fully reflect the historical landscape.
\item \textbf{Stealthy Activity Exclusion}: Extremely covert operations rarely yield samples suitable for inclusion.
\item \textbf{Dynamic Labeling}: APT attributions and naming evolve; labels may change over time.
\item \textbf{Sample-Type Predominance}: Executable files (especially Windows PE) dominate, potentially limiting coverage of certain platforms or payload types.
\item \textbf{Deduplication Scope}: The proposed deduplication is primarily optimized for binary executables and is not directly applicable to documents or scripts.
\item \textbf{Function-Clustering Dependency}: Function-level features rely on executable-binary analysis; non-executable or hard-to-analyze files have limited coverage.
\item \textbf{Feature Sensitivity of opcode\_cfg to Compilation and Disassembly}: The opcode\_cfg representation captures fine-grained opcodes on CFGs to prioritize high-precision matching; thus compiler/optimizer choices (e.g., inlining, instruction selection, register allocation, block layout), packing/obfuscation, and disassembly noise can alter opcodes or CFG shape, potentially reducing recall for semantically equivalent functions.
\end{itemize}

Researchers should consider these factors when using the dataset.

\section{Discussion and Future Work}

\label{sec:challenges}

This section discusses the key challenges encountered during the construction of APT-ClaritySet, the strategies adopted to address them, the advantages of the dataset, its potential applications, and future prospects, aiming to provide insights for related research.

\subsection{Challenges and Solutions}

Constructing a large-scale, high-quality APT dataset faces challenges such as attribution accuracy, alias system maintenance, deduplication limitations, efficiency of function feature extraction, and data source bias. This study effectively addressed these challenges through strategies including multi-level quality control (achieving a 96.43\% accuracy rate for organization labels), an iterable alias knowledge base (unifying approximately 11\% of inconsistent labels), an optimized graph-feature-based deduplication method, Faiss and GPU-accelerated function similarity computation (processing over 4.3 million functions), and multi-source data integration (covering 303 APT organizations, accounting for 32.44\% of known entities).

\subsection{Dataset Advantages and Applications}

The core advantages of APT-ClaritySet lie in its high-quality deduplication (eliminating 47.55\% of redundant samples) and standardized labels. The former significantly reduces storage and computational overhead and enhances the effectiveness of model training. The latter, through a unified APT organization alias system, supports cross-source intelligence integration and multi-dimensional analysis, such as research on APT activity trends and technological evolution paths. Furthermore, the high-quality function-level features, encompassing 324,538 function reuse clusters, provide a new perspective for APT code provenance, family evolution, and inter-organizational correlation analysis, contributing to the construction of a function-level ``fingerprint'' library and enabling inter-group code-sharing network analysis.

\subsection{APT Alias Mapping System Value}

The standardized alias system improves cross-institution intelligence integration (estimated $>65\%$) by unifying nomenclature, enabling automated processing and consistent situational awareness, and informing standardization efforts.

\subsection{Future Work}

Future work includes: 1) Methodological Improvements: Treat opcode\_cfg as a high-precision prefilter and integrate instruction/IR canonicalization, robust disassembly/lifting, and hybrid static–dynamic evidence; perform cross-compiler/optimization calibration and robustness evaluation under packing/obfuscation; further improve semantics-aware graph embeddings, principled thresholding, and real-time processing. 2) In-depth Research: Rigorous attribution with FRCs and the alias system (probabilistic/graph models), quantification of code reuse and provenance, and analysis of tooling across the kill chain. 3) Enhancing the Alias System: Community collaboration, intelligent alias management, and multilingual support.

\subsection{Community Contribution}

We will release core components (dataset, feature sets) and methodology at \url{https://github.com/APT-ClaritySet/Dataset}. Raw samples will be shared in disarmed form under a Data Use Agreement (DUA) for research use.

\section{Related Work}

\label{sec:related_work}

\subsection{Malware Data Collection \& Processing}

The malware-analysis pipeline typically comprises three phases: collection, processing, and annotation. Static analysis rapidly extracts features via disassembly and format parsing, but remains sensitive to packing/obfuscation and compiler diversity\cite{qiao2016automatically, sornil2013malware, moskovitch2008unknown, zhang2020ransomware, de2016lifted, zhang2019feature, pagani2018beyond, botacin2021understanding}. Dynamic analysis executes samples in instrumented environments to observe runtime behavior at the cost of scalability and potential evasions\cite{or2019dynamic, egele2014blanket, baldoni2017assisting, christodorescu2005semantics, brumley2007bitscope, moser2007exploring, ugarte2016rambo}. Hybrid approaches combine static robustness with dynamic fidelity to improve coverage\cite{jee2012general, jee2013shadowreplica, ming2016straighttaint, brumley2011bap, song2008bitblaze}.

Processing commonly extracts behavioral features (e.g., API/syscall sequences, network flows) and structural features (e.g., opcode n-grams, PE headers, CFG abstractions), leveraging entropy measures, machine learning, and multimodal fusion for representation and detection\cite{yakura2019neural, vasan2020imcfn, xiao2020malfcs, yuan2020byte, ghouti2020malware}. Annotation is constrained by expert effort and consistency; recent work explores automation via LLM-assisted labeling and active learning, coupled with multi-expert review to ensure quality\cite{alsulami2018behavioral, sanchez2025semantic, guo2024malosdf, lan2025trust, jelodar2025large, zhang2025automatically}. High-quality annotation is essential for reproducibility and for training models that generalize to emerging threats.

\subsection{Malware Similarity Detection \& Deduplication}

Similarity detection spans: (i) exact hashing, (ii) fuzzy hashing, and (iii) APT-oriented deduplication leveraging richer semantics. Exact hashes (e.g., MD5, SHA-1) identify byte-identical files but fail on polymorphic variants; collision concerns further motivate stronger digests such as SHA-256/SHA-3\cite{botacin2021understanding, cesare2010fast, or2019dynamic}. Fuzzy hashing (e.g., SSDeep, TLSH, ImpHash) measures approximate similarity in file structure or imports; threshold calibration is critical to balance recall and false positives\cite{kida2022nation, naik2021fuzzy}.

For APT datasets—often scarce yet highly polymorphic—hybrid deduplication combines fuzzy hashing with behavioral and structure-aware analysis, and employs learning-based clustering to capture families and near-duplicates\cite{zhang2025automatically, feng2016scalable}. Beyond file-level similarity, high-quality dataset construction also emphasizes sample diversity, temporal partitioning, and standardized attribution to support longitudinal studies and robust model training.
\section{Conclusion}

\label{sec:conclusion}

Addressing the scarcity of standardized APT malware benchmarks, this paper introduces APT-ClaritySet and a reproducible construction pipeline: 25,923 rigorously deduplicated and labeled samples mapped to 303 APT groups; a vendor-agnostic alias normalization system resolving cross-source naming inconsistencies; a deduplication scheme that couples CFG structural signatures with opcode semantics, effective under similar compilation settings but sensitive to compilation variations, reducing the analyzable set by 47.55\% while preserving behavioral diversity; and large-scale function-level code-reuse mining over 4.3M functions yielding 324,538 reuse clusters. Multi-dimensional validation (label validity, temporal consistency, cross-source agreement) indicates a group-label accuracy of 96.43\% at 95\% confidence. Together, the artifact and pipeline provide a standardized basis for measuring attack patterns, technique evolution, and attribution, and for building stronger detection and threat intelligence systems; to support responsible reproducibility, alias mappings, metadata, hashes, and construction scripts will be released under a research license, with access to binaries governed by legal and ethical controls.

{\appendices
\section{Deduplication Method Ablation Study}
\label{append:Ablation}
\subsection{Test Dataset Composition}

\noindent\textit{Ground-truth duplicate construction}. ``Duplicate'' denotes functionally equivalent binaries (not byte-identical). The test set (221 groups; 1181 samples) was created by applying minimal, functionality-preserving mutations—instruction-preserving edits (register renaming, NOP padding, basic-block reordering without changing control dependencies), non-functional constant/resource updates (strings, timestamps/metadata), and benign refactoring of non-critical helpers—while enforcing distinct cryptographic hashes. Equivalence was verified via sandbox behavior consistency (process tree, file/network I/O, exit status), stable external-interface checks, and sampled manual differential analysis.

\textbf{Dataset Scale and Distribution}:
\begin{itemize}
    \item \textbf{Duplicate Groups}: 221 (including 85 isolated groups)
    \item \textbf{Total Samples}: 1181
    \item \textbf{Group Size}: 2–69 per group (mean 8.06)
\end{itemize}

Two independent experts cross-validated the group-level similarity labels, yielding 96.8\% agreement.

\subsection{Experimental Configuration Parameters}

\begin{table}[htbp]
\centering
\caption{Ablation configuration matrix.}
\label{tab:config_matrix}
\begin{tabular}{clcccc}
\hline
\textbf{ID} & \textbf{Feature Rep.} & \textbf{Struct.} & \textbf{Semantic} & \textbf{Weight} & \textbf{Metric} \\
\hline
B1 & String & Centroid$^*$ & \ding{55} & \ding{55} & Cosine \\
B2 & String & $X,Y$ & \ding{55} & \ding{55} & Cosine \\
B3 & String & $X,Y,Z$ & \ding{55} & \ding{55} & Cosine \\
\hline
N1 & Numerical & Centroid$^*$ & \ding{55} & \ding{55} & Cosine \\
N2 & Numerical & $X,Y$ & \ding{55} & \ding{55} & Cosine \\
N3 & Numerical & $X,Y,Z$ & \ding{55} & \ding{55} & Cosine \\
\hline
O1 & String & $X,Y$ & \ding{51} & \ding{55} & Cosine \\
O2 & String & $X,Y,Z$ & \ding{51} & \ding{55} & Cosine \\
O3 & Numerical & $X,Y$ & \ding{51} & \ding{55} & Cosine \\
O4 & Numerical & $X,Y,Z$ & \ding{51} & \ding{55} & Cosine \\
\hline
W1 & Numerical & Centroid$^*$ & \ding{55} & \ding{51} & Cosine \\
W2 & Numerical & $X,Y$ & \ding{51} & \ding{51} & Cosine \\
W3 & Numerical & $X,Y,Z$ & \ding{51} & \ding{51} & Cosine \\
\hline
H1 & Numerical & $X,Y$ & \ding{51} & \ding{55} & Hybrid \\
H2 & Numerical & $X,Y,Z$ & \ding{51} & \ding{55} & Hybrid \\
H3 & Numerical & $X,Y$ & \ding{51} & \ding{51} & Hybrid \\
H4 & Numerical & $X,Y,Z$ & \ding{51} & \ding{51} & Hybrid \\
\hline
\multicolumn{6}{l}{\footnotesize $^*$ Chen et al.'s Centroid method~\cite{chen2014achieving}.}
\end{tabular}
\end{table}

\textbf{Key Parameter Settings}:
\begin{itemize}
    \item \textbf{Gaussian kernel}: $\gamma = 5$ for the structural component in the hybrid metric
    \item \textbf{WL propagation rounds}: $h = 1$
    \item \textbf{Semantic feature}: $W_{C_1}, \ldots, W_{C_9}$ denote CFG-path-weighted frequencies of nine opcode categories
    \item \textbf{CFG weighting}: function nodes (FCG) weighted by \texttt{CFG\_size}
\end{itemize}

\subsection{Ablation Experiment Results}
\label{sec:AblationResult}
Deduplication is cast as a pairwise binary classification task. For predicted duplicate pairs, we compute:
TP (true positives), FP (false positives), TN (true negatives), and FN (false negatives).
We sweep similarity thresholds $T \in \{0.99, 0.999, 0.9999, 0.99999, 0.999999\}$ and report Precision, Recall, and F1 at the F1-optimal $T$ for each configuration.

\begin{table}[htbp!]
\centering
\caption{Performance across 17 configurations (per-row best-F1 threshold).}
\label{tab:performance_results}
\begin{tabular}{cccccc}
\hline
\textbf{Config} & \textbf{Precision} & \textbf{Recall} & \textbf{F1} & \textbf{Threshold} & \textbf{Time (s)} \\
\hline
B1 & 0.9836 & 0.6798 & 0.8039 & 0.9999 & 18.27 \\
B2 & 0.9670 & 0.7131 & 0.8208 & 0.9999 & 18.73 \\
B3 & 0.9670 & 0.7131 & 0.8208 & 0.9999 & 18.71 \\
\hline
N1 & 0.8905 & 0.8242 & 0.8561 & 0.999999 & 22.89 \\
N2 & 0.8918 & 0.8380 & 0.8641 & 0.999999 & 22.57 \\
N3 & 0.8919 & 0.8379 & 0.8640 & 0.999999 & 23.02 \\
\hline
O1 & 0.9963 & 0.6703 & 0.8014 & 0.9999 & 18.92 \\
O2 & 0.9963 & 0.6703 & 0.8014 & 0.9999 & 20.38 \\
O3 & 0.9151 & 0.8263 & 0.8684 & 0.99999 & 26.52 \\
O4 & 0.8919 & 0.8379 & 0.8640 & 0.999999 & 24.92 \\
\hline
W1 & 0.9434 & 0.8852 & 0.9134 & 0.999999 & 29.72 \\
W2 & 0.8935 & 0.9187 & 0.9059 & 0.99999 & 30.59 \\
W3 & 0.8935 & 0.9187 & 0.9059 & 0.99999 & 29.65 \\
\hline
H1 & 0.8958 & 0.9139 & 0.9047 & 0.9999 & 24.18 \\
H2 & 0.9646 & 0.7846 & 0.8653 & 0.99999 & 24.26 \\
H3 & $\approx$1.0 & 1.0 & $\approx$1.0 & 0.9999 & 31.97 \\
H4 & 0.8609 & 0.9558 & 0.9059 & 0.9999 & 31.48 \\
\hline
\end{tabular}
\end{table}

\begin{table}[htbp]
\centering
\caption{Component-wise contributions.}
\label{tab:component_contributions}
\begin{tabular}{cccc}
\hline
\textbf{Component} & \textbf{Transition} & \textbf{$\Delta$F1} & \textbf{Overhead} \\
\hline
Numerical Rep. & B1$\to$N1 & +0.052 & 1.25$\times$ \\
Opcode Features & N2$\to$O3 & +0.004 & 1.18$\times$ \\
CFG Weighting & O3$\to$W2 & +0.038 & 1.15$\times$ \\
Hybrid Metrics & O3$\to$H1 & +0.036 & 0.91$\times$ \\
$Z$ Features & B2$\to$B3 & 0 & High* \\
\hline
\multicolumn{4}{l}{\footnotesize $^*$ Path explosion in complex functions during preprocessing.}
\end{tabular}
\end{table}

Computing $Z$ features can trigger path explosion on complex functions, incurring substantial preprocessing overhead.

% \begin{figure*}[htbp!]
% \centering
% \includegraphics[width=0.9\textwidth]{fig/single_component_contribution.pdf}
% \caption{Individual Component Comparison}
% \label{fig:single_component_contribution}
% \end{figure*}

\begin{figure}[htbp!]
\centering
\includegraphics[width=0.46\textwidth]{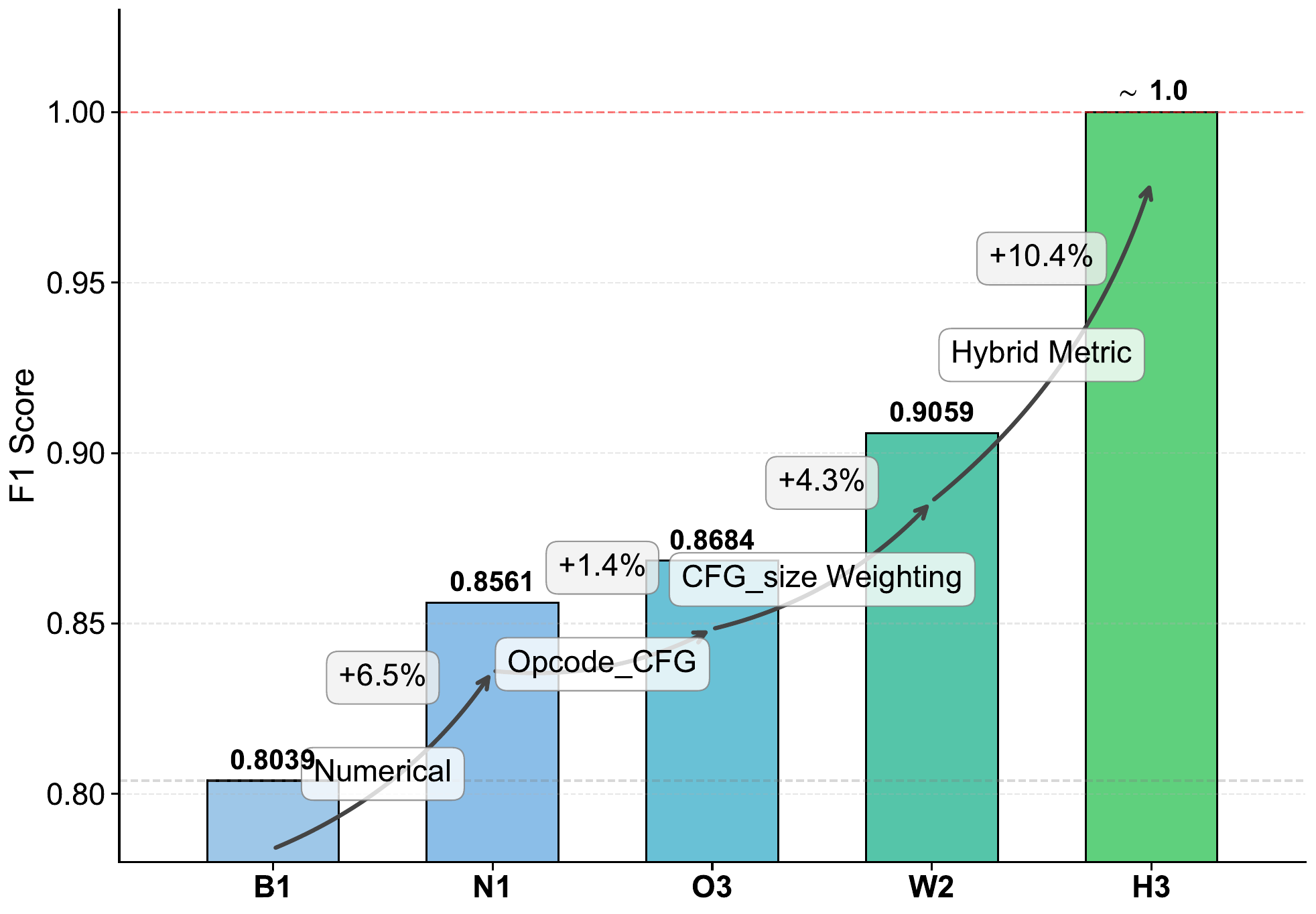}
\caption{Progressive enhancement path.}
\label{fig:progressive_enhancement}
\end{figure}

\begin{figure}[htbp!]
\centering
\includegraphics[width=0.46\textwidth]{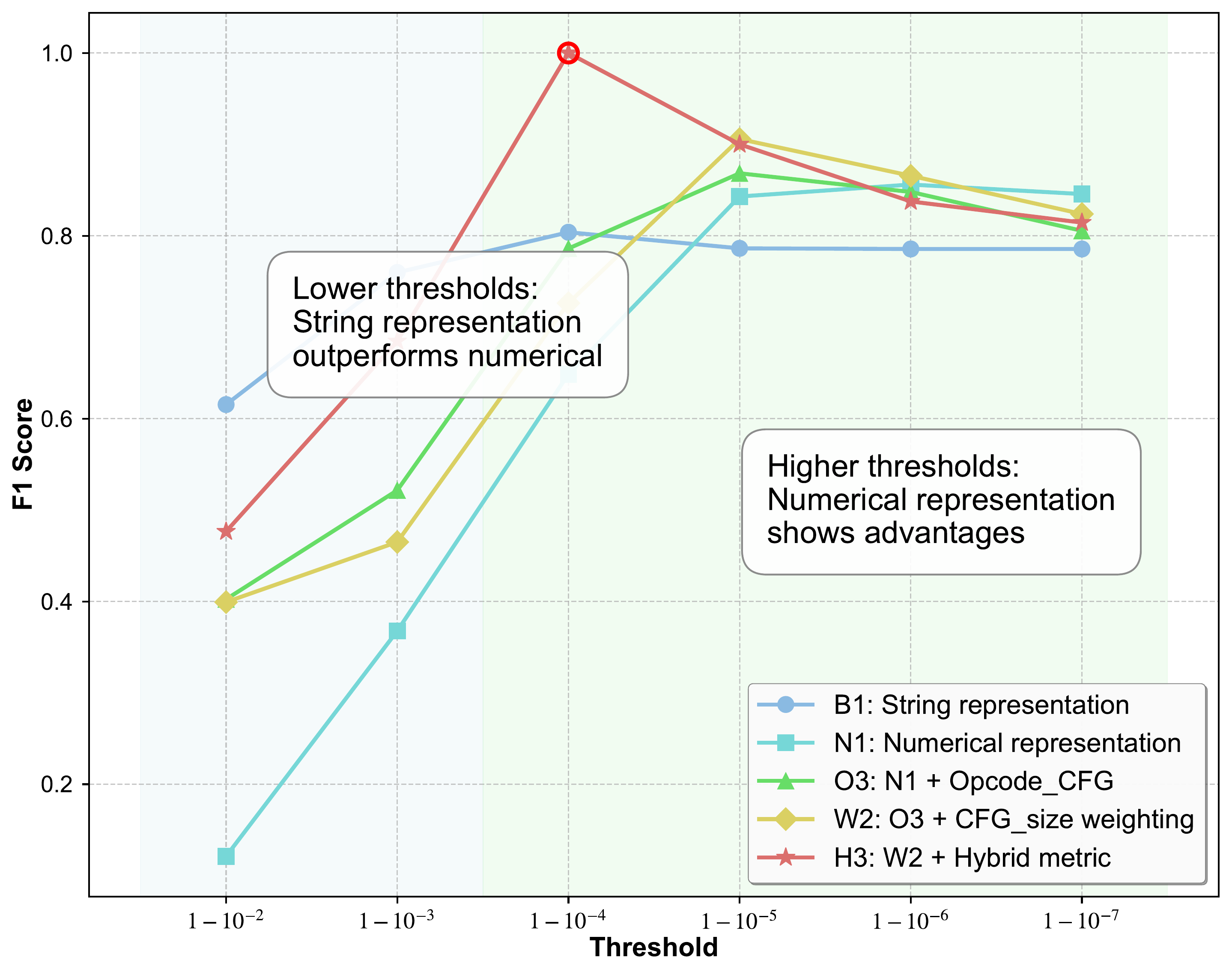}
\caption{F1 at different thresholds along the path.}
\label{fig:progressive_all_thresholds}
\end{figure}

Numerical representation and CFG weighting are the dominant contributors to F1 improvement, whereas opcode semantics and the hybrid metric offer smaller but consistent gains. As shown in Fig.~\ref{fig:progressive_enhancement}, performance improves progressively from B1 to H3. Fig.~\ref{fig:progressive_all_thresholds} further shows that string-based features are advantageous at low thresholds, while numerical representation dominates at high thresholds. We therefore adopt H3—structural $X,Y$ (without $Z$), opcode semantics, CFG weighting, and a hybrid similarity metric—as the default, achieving near-perfect F1 on the test set.

\section{Label Accuracy Assessment via Sampling}
\label{appendix:sampling}
This appendix evaluates the reliability of APT organization-level attribution labels in APT-ClaritySet using stratified random sampling and expert verification against primary threat intelligence sources.

\subsection{Sampling Design}
Given severe class imbalance across APT organizations (1--1{,}576 samples per organization), we adopted \textbf{stratified random sampling with near-optimal (Neyman) allocation}. In total, 1{,}906 samples (5.54\% of the dataset) were drawn according to:
\begin{itemize}
    \item Organizations with $\leq$5 samples: 100\% sampling
    \item Organizations with 6--20 samples: 50\% sampling ($\geq$5 samples)
    \item Organizations with 21--100 samples: 30\% sampling ($\geq$10 samples)
    \item Organizations with 101--500 samples: 15\% sampling ($\geq$30 samples)
    \item Organizations with $>$500 samples: 8\% sampling (75--120 samples)
\end{itemize}
This design ensured coverage of all 305 APT organizations and controlled variance via internal stratification by discovery time (before 2015, 2015--2019, and $\geq$2020).

\subsection{Expert Source Verification}
For each sampled label, experts traced primary threat intelligence sources to verify: (i) whether the source explicitly attributed the sample to the labeled APT organization; and (ii) whether the recorded hash matched the referenced artifact. Missing or ambiguous cases were supplemented with secondary sources and cross-validated across multiple intelligence databases.

\subsection{Results}
Among the 1{,}906 sampled instances:
\begin{itemize}
    \item Correct: 1{,}838 (96.43\%)
    \item Suspicious: 40 (2.10\%)
    \item Incorrect: 11 (0.58\%)
    \item Unknown: 17 (0.89\%)
\end{itemize}
Among 1{,}878 cases with locatable provenance, 1{,}503 (80.0\%) were confirmed directly from first-party threat intelligence databases, while 375 (20.0\%) required external corroboration. At the 95\% confidence level, the estimated accuracy of APT organizational attribution labels is \textbf{96.43\%} with confidence interval [95.49\%, 97.21\%].

}

\bibliography{paper1}{}
\bibliographystyle{IEEEtran}

\vfill

\end{document}